\documentstyle[12pt,epsf]{article}

\def\Title#1{\begin{center} {\Large #1 } \end{center}}
\def\Author#1{\begin{center}{ \sc #1} \end{center}}
\def\Address#1{\begin{center}{ \it #1} \end{center}}
\def\andauth{\begin{center}{and} \end{center}}

\newenvironment{Abstract}{\begin{quotation} \begin{center}
                       ABSTRACT
     \end{center}\bigskip  }{\end{quotation}}
\def\beq{\begin{equation}}
\def\eeq#1{\label{#1}\end{equation}}
\def\eeqn{\end{equation}}
\def\beqa{\begin{eqnarray}}
\def\eeqa#1{\label{#1}\end{eqnarray}}
\def\eeqan{\end{eqnarray}}

\def\Acknowledgements{\bigskip  \bigskip \begin{center} \begin{large}
             \bf ACKNOWLEDGEMENTS \end{large}\end{center}}
\def\bar#1{\overline{#1}}
\def\Dslash{\not{\hbox{\kern-4pt $D$}}}

\def\OCIP{Ottawa-Carleton Institute for Physics \\
     Department of Physics, Carleton University, Ottawa CANADA, K1S 5B6}
\def\UQAM{D\'epartement de  Physique, Universit\'e
du Qu\'ebec \`a Montr\'eal \\
C.P. 8888, Succ. Centre-Ville, Montr\'eal, Qu\'ebec, Canada, H3C 3P8}
\begin{document}
\mbox{}\hfill OCIP/C-96-2\\
\mbox{}\hfill UQAM-PHE-96-08\\
\Title{MEASUREMENT OF THE $WW\gamma$ and $WWZ$ COUPLINGS \\
IN THE PROCESS $e^+e^- \to \ell \nu q\bar{q}'$}
\bigskip
\Author{Mikul\'{a}\v{s} Gintner, Stephen Godfrey}
\medskip
\Address{\OCIP}
\medskip
\andauth
\medskip
\Author{Gilles Couture}
\Address{\UQAM}
\bigskip
\begin{Abstract}
We studied the sensitivity of the process
$e^+e^-\to \ell \nu q\bar{q}'$ to anomalous
trilinear gauge boson couplings of the $WW\gamma$
and $WWZ$ vertices at the center of mass energies $\sqrt{s}=500$~GeV
and 1~TeV. The bounds for the couplings we obtained result from an
analysis of a five dimensional angular differential cross section.
In our calculations we included all tree level Feynman diagrams
contributing to the final state as well as the finite widths of
the vector bosons. Both unpolarized and polarized beams were considered.
We found that the 500~GeV measurements will be at the level of loop
contributions to the couplings and may show hints of new physics while
the 1~TeV  should be sensitive to new physics at the loop level.
We also explored $\ell\nu q\bar{q}$ final states off the $W$
resonance and found that useful information could be extracted
from this region of phase space.
\end{Abstract}
\bigskip

\def\thefootnote{\fnsymbol{footnote}}
\setcounter{footnote}{0}
\section{Introduction}

One of the important goals of the Next
Linear Collider (NLC) is to
make precision measurements of $W$ boson properties including
$W$-boson interactions with the photon and $Z^0$.
The measurements of the trilinear gauge boson vertices (TGV's)
provide a stringent test of the gauge structure of the standard
model\cite{tgvreviews,aihara95}.
The current measurement of these couplings
performed by the CDF and D0 collaborations \cite{tevatron}
are rather weak. Although
the measurements at the Large Hadron Collider should
improve these limits considerably it is expected that
measurements at high energy $e^+e^-$ colliders will
surpass those at the hadron colliders.

Probably the most useful of the $e^+e^-\to W^+W^-$ channels
for these measurements is $e^+e^- \to \ell
\nu q \bar{q}'$.
With only one unobserved neutrino this channel has several advantages:
it can be fully reconstructed
using the constraint of the initial beam energies,
the  $W^+$ and $W^-$ can be discriminated using lepton charge identification,
it does not have the QCD backgrounds that plague the fully hadronic
decay modes, and it offers much higher statistics than the fully
leptonic modes.

In this paper we examine the sensitivity of this channel
to anomalous $WW\gamma$ and $WWZ^0$ gauge boson couplings.
We study this process for the centre of mass energies
 $\sqrt{s}=500$ and
1000~GeV appropriate to the NLC.
In our calculations we included all tree level diagrams to the four
fermion final states using helicity amplitude techniques
and we considered the finite widths of the vector bosons.
To determine the sensitivity of the channel to anomalous gauge boson
couplings we examined numerous distributions \cite{gigocu95}. We
also looked at the question of how important 
the non-resonant diagrams are in these calculations and how much
information about the TGV's can be found in the off-resonance
production. Using helicity
amplitudes we are able to study the usefulness of initial state
polarization in extracting the TGV's.

In the next section we discuss the effective Lagrangian parametrization
and make introductory remarks on our calculations.
In section 3 we present and discuss our results.
We summarize our conclusions in section 4.

\section{Calculations}

\subsection{Parametrization}

In order to describe TGV's we use the most general
parametrization possible that respects Lorentz invariance,
electromagnetic gauge invariance and $CP$ invariance
\cite{hagiwara87,gaemers79,miscvertex} .
These constraints leave us with five
free independent parameters for the $WW\gamma$ and $WWZ$ vertices
and the effective interaction lagrangian is given by
\begin{equation}
{\cal L}_{WWV} =  - ig_V \left\{ { g_1^V (W^+_{\mu\nu}W^{-\mu} -
W^{+\mu}  W_{\mu\nu} ) V^\nu
+ \kappa_V W^+_\mu W^-_\nu V^{\mu\nu}
+ {{\lambda_V}\over{M_W^2}} W^+_{\lambda\mu}W^{-\mu}_\nu V^{\nu\lambda}
}\right\}
\end{equation}
where the subscript $V$ denotes either a photon or a $Z^0$,
$V^\mu$ and $W^\mu$ represents  the photon or $Z^0$ and $W^-$
fields respectively,
$W_{\mu\nu}=\partial_\mu W_\nu-\partial_\nu W_\mu$ and
$V_{\mu\nu}=\partial_\mu V_\nu-\partial_\nu V_\mu$
and $M_W$ is the $W$ boson mass.  ($g_1^\gamma$ is constrained by
electromagnetic gauge invariance to be equal to 1.)
At tree level the standard model requires $g_1^Z=\kappa_V=1$ and
$\lambda_V=0$.

This approach has become the standard parametrization used
in phenomenology making the comparison
of the sensitivity of different measurements
to the TGV's straightforward. Later in the paper we show
bounds that could be imposed on these five parameters
measuring the $e^+e^- \to \ell \nu q \bar{q}'$ process
at NLC machines.
We also calculate the sensitivity
to the $L_{9L}$ and $L_{9R}$ parameters \cite{gigocu95} 
that appear in
the chiral Lagrangian parametrization of
the effective TGV's.

\subsection{Initial remarks and conditions}

There are
10 (20) diagrams contributing to the $\mu^\pm \nu_\mu q \bar{q}'$ 
($e^\pm \nu_e q\bar{q}'$)
final state \cite{gigocu95}.
Of these there are two (three) diagrams which contain TGV's in the
muon (electron) mode of the reaction.
To include finite width effects we used vector boson propagators
of the form $(s-M_V^2 + i\Gamma_V M_V)^{-1}$ which yields
a gauge invariant result.
We used the CALKUL helicity amplitude technique \cite{calkul} to
obtain expressions for the matrix elements and
performed the phase space integration using  Monte Carlo
techniques.
To obtain numerical results we used the values $\alpha=1/128$,
$\sin^2\theta=0.23$, $M_Z=91.187$ GeV, $\Gamma_Z=2.49$ GeV,
$M_W=80.22$ GeV, and
$\Gamma_W=2.08$ GeV.  In our results we included two generations of
quarks and took them to be massless.
In order to take into account finite detector acceptance we require
that the lepton and quarks are at least
10 degrees away from the beam and have at least 10~GeV energy 
(with exception of Table \ref{xstab}).

\section{Results}

\subsection{Total cross section}

We calculate total cross sections 
as a function of $\sqrt{s}$
for both lepton modes \cite{gigocu95}. The values for energies 
of 175~GeV, 500~GeV, and 1~TeV
are given in Table \ref{xstab}.
In general the electron mode has a larger cross-section than the
muon mode.  The difference is small at 175~GeV but becomes
increasingly larger at higher energy as the t-channel photon
exchange  becomes increasingly important, reaching a factor of 5
at 1~TeV.


Imposing cuts on the invariant masses of the $\ell\nu$ and $q\bar{q}'$
pairs we can assess contributions of the non-resonant diagrams
to the total cross section.
For the muon mode the invariant mass cuts reduce the
cross-section by 10\% to 20\% depending on $\sqrt{s}$ and
irrespective of whether the cut is on $M_{\ell\nu}$ or $M_{q\bar{q}}$.
The relatively small effect of these cuts
verifies the dominance of the resonant diagrams on the total cross
section.
For the electron mode the cross section with the cut on $M_{q\bar{q}'}$
is significantly larger than the cross section with the cut
on $M_{e\nu}$. This can be attributed to poles in the single $W$
production
t-channel photon exchange diagrams.  With appropriate
kinematic cuts this can be used to study single-$W$ production.

Despite the relative smallness of the
off-resonance contributions to the muon mode they still contribute
up to 30\% of the cross section at 1~TeV.  Clearly, they must be
properly included when making high precision tests of standard model
processes.  For the electron mode they are even more important and
are interesting in the context of single $W$ production.

\begin{table}
\caption{\it Cross-sections for $e^+e^- \to \mu^+\nu_\mu q\bar{q}'$ and
$e^+e^- \to e^+\nu_e q\bar{q}'$ including cuts on the invariant
masses of the outgoing fermion pairs, $M_{\ell\nu}$ and $M_{q\bar{q}'}$.
A $10^o$ cut away from the beam is imposed on charged final state
fermions and no cut on their energy.
The cross-sections are given in pb.}
\begin{center}
\begin{tabular}{|c|c|c|c|c|c|} \hline
$\sqrt{s}$ & $\ell$ & no cut & $|M_{q\bar{q}}-M_W|<5$~GeV &
        $| M_{\ell\nu}-M_W|<5$~GeV & both cuts \\
(GeV) &   &  &  &  &  \\
\hline\hline
175 & $\mu$ & 1.10 & 1.00 & 1.00 & 0.91 \\
        & $e$ & 1.15 & 1.04 & 1.01 & 0.91 \\
\hline
500 & $\mu$ & 0.39 & 0.34 & 0.34 & 0.29 \\
        & $e$ & 0.62 & 0.53 & 0.34 & 0.29 \\
\hline
1000 & $\mu$ & 0.077 & 0.063 & 0.064 & 0.052 \\
        & $e$ & 0.44 & 0.39 & 0.064 & 0.052 \\
\hline
\end{tabular}
\end{center}
\label{xstab}\end{table}

\subsection{Distributions}

We examined numerous distributions and applied various kinematic
cuts to identify the observables and the regions of phase space
that are the most sensitive to anomalous couplings.
Details can be found in ref. \cite{gigocu95}.
Here we summarize the most interesting points and our conclusions.

Any disruption of the delicate gauge theory cancellations leads to
large changes to the Standard model results.  For
$W_L$ production amplitudes the enhancements can be a factor of
$(s/M_W^2)$.
Because it is the longitudinal $W$ production which is the most
sensitive to anomalous couplings, and because the cross section is
dominated by transverse $W$ production it is crucial to disentangle
the $W_L$ from the $W_T$ {\it background}.
The angular distributions of the $W$ decay products provide the most
convenient tool for doing so.

To investigate the sensitivity of
the $e^+e^- \to \ell^\pm \nu_\ell q \bar{q}'$ process to
the anomalous couplings we used
a combined distribution of five angular observables
\cite{gigocu95,barklow92,sekulin} :
$\Theta$, $\theta_{qq}$, $\phi_{qq}$, $\theta_{\ell\nu}$, and
$\phi_{\ell\nu}$, where
$\Theta$ is the $W^-$
scattering angle with respect to the initial $e^+$ direction,
$\theta_{qq}$ is the polar decay angle of the $q$ in the $W^-$ rest
frame using the $W^-$ direction as the quantization axis,
$\phi_{qq}$ is the azimuthal decay angle of the $q$ in the $W^-$ rest
frame, and $\theta_{\ell\nu}$ and $\phi_{\ell\nu}$ are the analogous
angles for the lepton in the $W^+$ rest frame.
There is an ambiguity in measurement of the $q\bar{q}$ 
pair observables since we cannot tell which hadronic jet
corresponds to the quark and which to the antiquark.  We therefore
include both possibilities in our analysis.
Each of the variables was divided into four bins so that
the entire phase space was divided into $4^5 = 1024$ bins.
To extract the information about the TGV's we used
the maximum likelihood method using Poisson statistics
because with this many bins some will not
be very populated with events. 
The results we obtained are based solely on
the statistical errors using
the integrated lumininosity we assumed for the various cases.

\begin{figure}
\begin{center}
\leavevmode
{\epsfxsize=2.00truein \epsfbox{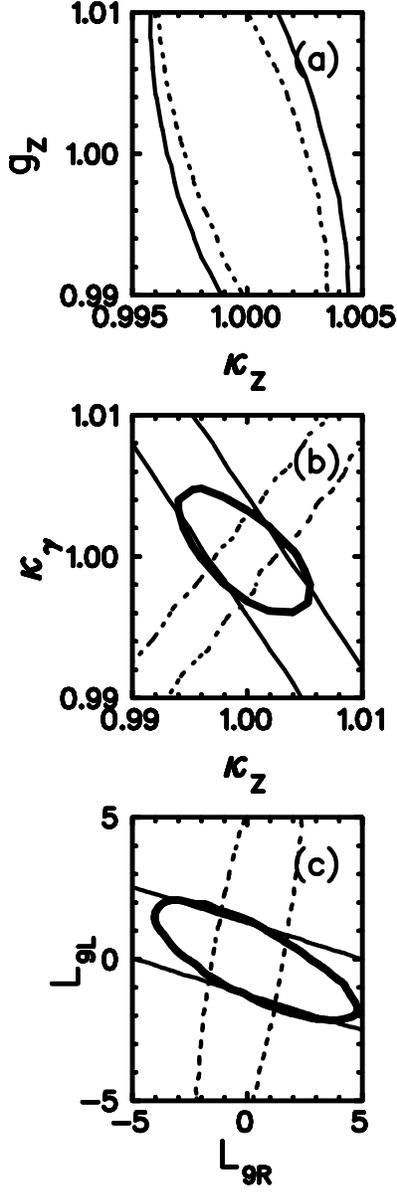}}
\end{center}
 \caption[foo]{95\% C.L. exclusion contours for sensitivity to 
 anomalous couplings for unpolarized (the heavy solid line) and
polarized initial state electrons with $\sqrt{s}=500$~GeV,
combining all four lepton charge states.
The solid curves are for $e^-_{L}e^+$, the dashed curves for
$e^-_{R}e^+$. The integrated luminosity is $L=25$~fb$^{-1}$ 
per polarization, for unpolarized electrons $L=50$~fb$^{-1}$.}
\label{500fig}
\end{figure}

A thorough analysis of gauge boson couplings would allow all five
parameters in the Lagrangian to vary simultaneously to take into
account cancellations (and correlations) among the various contributions.
This approach is  impractical, however, due to the large
amount of computer time that would be required to search the parameter
space.  Instead we found 2-dimensional {\it exclusion}
 contours for a selection
of parameter pairs to give a sense of the correlations.
Because of limited space, here we show only limits obtained
for individual parameters when all the other
anomalous parameters
are kept fixed at their Standard model values (Table \ref{limitstab}).
In Figure \ref{500fig} we show a few representative 
contours for $\sqrt{s} = 500$~GeV.
More contours can be found in \cite{gigocu95}.

\begin{table}
\caption{\it Sensitivities to anomalous couplings for the various
parameters varying one parameter at a time.  The combined values
are obtained
by combining the four lepton modes ($e^-$, $e^+$, $\mu^-$, and
$\mu^+$) and two generations of light quarks ($ud$, $cs$).
The results are  95\% confidence level limits.}
\begin{center}
\begin{tabular}{|l|c|c|c|c|c|c|c|}
\hline
mode &  $\delta g_1^Z$ & $\delta\kappa_Z$ &
        $\delta\kappa_\gamma$ & $\lambda_Z$ & $\lambda_\gamma$ &
        $L_{9L}$ & $L_{9R}$ \\
\hline\hline
\multicolumn{8}{|c|}{$\sqrt{s}=500$~GeV, L=50~fb$^{-1}$,
$|M_{\ell\nu (q\bar{q})}- M_W|<10$~GeV }\\
\hline\hline
$\mu$  & $\pm 0.020$ &
        $\pm 0.007$ & $\pm 0.005$ & $\pm 0.005$
        & $\pm 0.006$ 
        & $^{+2.2}_{-2.1}$ & $^{+4.6}_{-4.2}$\\
$e$ & $^{+0.019}_{-0.020}$ &
        $\pm 0.007$ & $\pm 0.005$ & $\pm 0.005$
        & $\pm 0.006$ 
        & $^{+2.2}_{-2.1}$ & $^{+4.6}_{-4.2}$ \\
combined & $\pm 0.0095$ &
        $\pm 0.0035$ & $\pm 0.0025$ & $\pm 0.0025$
        & $\pm 0.0025$ 
        & $^{+1.1}_{-1.0}$ & $^{+2.2}_{-2.1}$ \\
\hline\hline
\multicolumn{8}{|c|}{$\sqrt{s}=1$~TeV, L=200~fb$^{-1}$,
$|M_{\ell\nu (q\bar{q})}- M_W|<10$~GeV }\\
\hline\hline
$\mu$ & $\pm 0.01$ &
        $\pm 0.002$ & $\pm 0.001$ & $\pm 0.002$
        & $\pm 0.002$ 
        & $^{+0.61}_{-0.62}$ & $^{+1.3}_{-1.1}$ \\
$e$ & $\pm 0.01$ &
        $\pm 0.002$ & $\pm 0.001$ & $\pm 0.002$
        & $\pm 0.002$ 
        & $^{+0.61}_{-0.62}$ & $^{+1.3}_{-1.1}$ \\
combined & $\pm 0.0054$ &
        $\pm 0.001$ & $\pm 0.0006$ & $\pm 0.0008$
        & $\pm 0.0008$ 
        & $\pm 0.28$ & $^{+0.62}_{-0.56}$ \\
\hline
\end{tabular}
\end{center}
\label{limitstab}\end{table}

It was pointed out that the angular distributions
with different initial electron polarizations have
different dependences on anomalous couplings \cite{gigocu95,pankov95}.
We calculated contours for $e_L^- e^+$ and $e_R^- e^+$ initial states
\cite{gigocu95}.
An important difference
between the two polarizations is that
the cross-section with left handed electrons is about an order of
magnitude larger than for right handed electrons.  At the same time
the right-handed cross-section is significantly more sensitive to anomalous
couplings than the left-handed cross-section.  There is therefore
a tradeoff between sensitivity and statistics so that in most cases
the single parameter
bounds obtainable for the two polarizations are comparable.
In general, the unpolarized contours are aligned along the $e_L^-$
contours, which is not too suprising considering that
$\sigma (e^-_L)$ dominates over
the right-handed contribution in the unpolarized cross-section.
For some combinations of
the parameters the different polarizations give much different
correlations which could be useful for disentangling
the nature of anomalous TGV's if deviations were observed.

\subsection{Off-resonance production}

In this section we explore the information potential available from
$\ell\nu q\bar{q}$ final states off the $W$ resonance.
We calculated invariant mass distributions
($M_{\mu \nu}$, $M_{e\nu}$, and
$M_{q\bar{q}'}$) of final state fermions with
various polarizations of the initial electron
\cite{gigocu95}. Varying TGV's parameters and looking at
the response of these distributions we found that
there is considerable sensitivity to anomalous gauge
boson couplings even when the fermion pairs do not 
originate from real $W$ production.

To quantitatively demonstrate that there is considerable
information in the non-resonant production
we find limits for the TGV's parameters based
on the total cross section while imposing the cuts
$M_{ff'}<M_W-15$~GeV and $M_{ff'}>M_W+15$~GeV. Here, the
$M_{ff'}$ is the invariant mass of the final state fermion pairs and
$ff'$ stands for either $\ell \nu$ or $q\bar{q}$.  These give rise to a
large number of possibilities so we only present the
``best'' case when the four possible final states are combined for
each energy. The most significant bounds are presented in 
Table \ref{offshelltab}.
We warn that our analysis is not a rigorous one.
 In particular we do not consider
possible backgrounds to non-resonant events and do not make any
effort to optimize our cuts to enhance deviations from SM results.

\begin{table}
\caption{\it Sensitivities to anomalous couplings based on off-resonance
cross sections varying one parameter at a time.  The values are obtained
by combining the four lepton modes ($e^-$, $e^+$, $\mu^-$, and
$\mu^+$) and two generations of light quarks ($ud$, $cs$).
The results
are  95\% confidence level limits. A dash signifies that the bound
is signficantly weaker than the others.}
\begin{center}
\begin{tabular}{|c|c|c|c|c|c|c|c|c|}
\hline
Initial State & cut & $\delta g_{1Z}$
        & $\delta\kappa_Z$ &
        $\delta\kappa_\gamma$ & $\lambda_Z$ & $\lambda_\gamma$ &
        $L_{9L}$ & $L_{9R}$ \\
\hline\hline
\multicolumn{9}{|c|}{$\sqrt{s}=500$~GeV, L=50~fb$^{-1}$ }\\
\hline\hline
$e_L^-$ & $M_{\ell\nu}>M_W +15$~GeV & $_{-0.18}^{+0.13}$
        & $_{-0.074}^{+0.056}$ & $_{-0.012}^{+0.012}$ &
$_{-0.034}^{+0.026}$ & $_{-0.041}^{+0.026}$ 
& $_{-7.2}^{+6.9}$ & $_{-9.2}^{+8.9}$ \\
$e_R^-$ & $M_{\ell\nu}>M_W +15$~GeV &
$_{-0.24}^{+0.08}$ & $_{-0.10}^{+0.03}$ & $_{-0.018}^{---}$ &
$_{-0.033}^{+0.023}$ & $_{-0.023}^{+0.034}$ 
& $_{---}^{---}$ & $_{---}^{---}$ \\
$e^-$ & $M_{\ell\nu}>M_W +15$~GeV &
$_{-0.21}^{+0.15}$ & $_{-0.10}^{+0.07}$ & $_{-0.017}^{+0.016}$ &
$_{-0.038}^{+0.031}$ & $_{-0.043}^{+0.033}$ 
& $_{-10}^{+10}$ & $_{-13}^{+13}$ \\
\hline\hline
\multicolumn{9}{|c|}{$\sqrt{s}=1$~TeV, L=200~fb$^{-1}$}\\
\hline\hline
$e_L^-$ & $M_{\ell\nu}>M_W +15$~GeV &
$_{-0.085}^{+0.038}$ & $_{-0.014}^{+0.013}$ & $_{-0.005}^{+0.005}$ &
$_{-0.002}^{+0.002}$ & $_{-0.003}^{+0.003}$ 
& $_{-2.5}^{+2.4}$ & $_{-3.9}^{+3.8}$  \\
$e_R^-$ & $M_{\ell\nu}>M_W +15$~GeV &
$_{-0.102}^{+0.062}$ & $_{---}^{+0.010}$ &
$_{-0.010}^{---}$ & $_{-0.007}^{+0.007}$ & $_{-0.007}^{+0.007}$ 
& $_{---}^{---}$ & $_{-5.9}^{---}$ \\
$e^-$ & $M_{\ell\nu}>M_W +15$~GeV & $_{-0.096}^{+0.047}$
        & $_{-0.021}^{+0.018}$ & $_{-0.007}^{+0.007}$ &
$_{-0.003}^{+0.003}$ & $_{-0.003}^{+0.003}$ 
& $_{-3.5}^{+3.4}$ & $_{-5.6}^{+5.4}$ \\
\hline
\end{tabular}
\end{center}
\label{offshelltab}\end{table}

From the above results it is clear that, although the constraints
that could be obtained from off-resonance production are not as
tight as those obtained from on-shell $W$ production,  there is
nevertheless considerable information contained in these events.
It appears to us that the method that makes optimal use of each
event is to calculate the probability of each event, irrespective of
where it appears in phase space, and compute a likelihood function
for the combined probabilities.  The only cuts that
should be included are those that represent detector acceptance and
that are introduced to eliminate backgrounds.

\section{Conclusions}

We performed a detailed analysis of the measurement of tri-linear
gauge boson couplings in the process $e^+e^- \to \ell^\pm \nu q
\bar{q}$.
To gauge the sensitivity of this process to anomalous gauge boson
couplings we used the $W$ decay distributions as a polarimeter to
distinguish the longitudinal $W$ modes, which are more sensitive to
anomalous coupings, from the transverse modes.  We implemented this
through the use of a quintic differential cross section, with each
angular variable divided into 4 bins, and then calculating the
likelihood function of non-standard model couplings as compared to
the standard model.
The 500~GeV NLC measurements are sensitive enough that they should
be sensitive to loop contributions to the TGV's while the 1~TeV will
be able to measure such effects.  

We studied the sensitivity of the off-mass shell cross sections
to anomalous couplings by imposing kinematic cuts on the invariant
mass distributions of the outgoing fermion pairs.
A cursory analysis found that the off-resonance cross section is
relatively sensitive to anomalous couplings and that useful
information could be extracted from this region of phase space.

The optimal strategy to maximize the information contained in each
event is to construct a likelihood function based on the four vector
of each of the outgoing fermions on an event by event basis, putting
them through a realistic detector simulation.  This would make the
best use of the information whether it be on the $W$ resonance or
not.  Kinematic cuts should only be introduced to reduce backgrounds.
Since the precision of these measurements is beyond the level of
loop induced radiative corrections it is crucial that radiative
corrections are well understood and included in event generators
used in the study of these processes.

  \Acknowledgements

The authors benefited greatly from many helpful conversations,
communications, and suggestions  during the course of this work
with Tim Barklow, Genevieve B\'elanger,
Pat Kalyniak, Dean Karlen, and Paul Madsen.
This research was supported in part by the Natural Sciences and
Engineering Research Council of Canada and Les Fonds FCAR du Quebec.



\end{document}